\documentclass{article}

\usepackage{icrctc07}

\def\be{\begin{equation}}
\def\ee{\end{equation}}
\def\ba{\begin{eqnarray}}
\def\ea{\end{eqnarray}}

\def\lsim{\raise0.3ex\hbox{$\;<$\kern-0.75em\raise-1.1ex\hbox{$\sim\;$}}}
\def\gsim{\raise0.3ex\hbox{$\;>$\kern-0.75em\raise-1.1ex\hbox{$\sim\;$}}}

\def\theta{\vartheta}

\title{Constraints on secondary 10-100 EeV gamma ray flux in the
minimal bottom-up model of Ultra High Energy Cosmic Rays.}
\shorttitle{Constraints on UHE gamma-rays}
\authors{O.Kalashev$^1$, G.Gelmini$^2$ and D.V.Semikoz$^{1,3}$}
\shortauthors{G.Gelmini, O.Kalashev and D.V.Semikoz}
\afiliations{
$^1$INR RAS, 60th October Anniversary prospect 7a, 117312 Moscow, Russia\\
$^2$ UCLA, Los Angeles, CA 90095-1547, USA\\
$^3$APC, 10 rue Alice Domon and Leonie Duquet, Paris 75205, France
} 
\email{e-mail: kalashev@ms2.inr.ac.ru}

\abstract{
In a recently proposed model the cosmic rays
spectrum at energies above EeV can be fitted with a minimal number of
unknown parameters assuming that the extragalactic cosmic rays are
only protons with a power law source spectrum~\cite{dip}. Within this minimal
model, after fitting the observed HiRes spectrum with four parameters
(proton injection spectrum power law index, maximum energy, minimum
distance to sources and evolution parameter) we compute the flux of
ultra-high energy photons due to photon-pion production and e+e- pair
production by protons for several radio background models and a range
of average extragalactic magnetic fields.
}

\begin{document}

\maketitle

The  ``ankle"   in the Ultra-high Energy Cosmic Ray (UHECR) spectrum can be interpreted as an absorption ``dip" at energies
$E=3-10$~EeV~\cite{dip}, due to the
propagation of extragalactic protons over large distances in the
cosmic microwave background (CMB)~\cite{dip-old}. 
 This would agree with
the indication of a transition from heavy to light primary nuclei
observed by the HiRes collaboration at energies close to 5$\times
10^{17}$ eV~\cite{chem_HiRes}.
 In this case the UHECR  HiRes
spectrum~\cite{HiRes}, in which the GZK
cutoff~\cite{gzk} 
is present,   can be fitted with a minimal number of
unknown parameters assuming the extragalactic cosmic rays are only
protons with a power law source spectrum $\sim E^{-\alpha}$ with
$\alpha\simeq 2.6$~\cite{dip}.  This is a minimal model of UHECR.

The GZK process produces pions. 
From the
decay of $\pi^0$ we obtain photons,  which we call ``GZK photons".
Previously we studied in detail the GZK photon flux dependence on
different unknown parameters of the source spectrum and
distribution and the intervening cosmological
backgrounds~\cite{gzk_photon}. Below we breafly discuss the perspectives for
photon detection in the minimal UHECR model (for more details see Ref.~\cite{min_photon}).

We use a numerical code developed in Ref.~\cite{kks1999,gzk_photon} to
compute the flux of GZK photons produced by a homogeneous
distribution of sources emitting originally only protons. This is the
same numerical code as in Ref.~\cite{gzk_photon}, with a few
modifications described in details in ref.~\cite{min_photon}.

As it is usual, we take the spectrum of an individual UHECR source  to
be of  the form:
\be  F(E) = f E^{-\alpha} ~~\Theta (E_{\rm max} -E)
 \label{proton_flux}
\ee
where $f$ provides the flux normalization, $\alpha$ is  the spectral  index and
$E_{\rm max}$ is the maximum energy to which protons can be
accelerated at the source.
The source density  is
defined by
\be n(z) = n_0 (1+z)^{3+m}  \Theta (z_{\max}-z) \theta (z-z_{\min}) \,,
\label{sources}
\ee
where $m$  parameterizes the source density evolution ($m=0$ corresponds to non-evolving sources with constant density
per comoving volume) and $z_{\min}$ and $z_{\max}$ are respectively
the redshifts of the closest and most distant sources.  Sources with
$z>2$ have a negligible contribution  to the UHECR flux
above $10^{18}$~eV. The value of $z_{\min}$ is connected to the
density of sources and influences strongly the shape of  the ``bump"
produced by the pile-up of protons which loose energy  in the GZK
cutoff and the strength of the GZK
suppression~\cite{kst2003,blasi2003,sources2004}.  Here we
fix $z_{\rm max} =3$ and consider three values for $z_{\rm min}$,
namely  0,  0.005 and  0.01 in Eq.~(\ref{sources}).

The main energy loss mechanism for photons with $E>10^{19}$~eV is
pair production on the radio background and cascade electron and positrons 
losses in the Extra  Galactic Magnetic Fields (EGMF).
Here we assume either the  minimal
intervening  radio   background of Clark {\it et al.}~\cite{clark}) 
and EGMF $B=10^{-11}$ G or  the largest radio background of Protheroe and
Biermann~\cite{PB}) and  EGMF $B=10^{-9}$ G, and many different
source models.

We consider many different proton spectra resulting from changing   the
slope $\alpha$ and the maximum energy $E_{\max}$  in
Eq.~\ref{proton_flux} within the ranges $2.3 \leq \alpha\leq 2.8$ and
$1.6 \times 10^{20} {\rm eV}\leq E_{\max}\leq 1.28 \times 10^{21}$~eV
and the source evolution parameter $m$ in Eq.(\ref{sources})  within
the range $-2 \leq m \leq 3$. 
 We  fit the observed spectrum UHECR~\cite{HiRes} at energies $E \ge 2$ EeV with these models,
  which requires a steaply falling source proton spectra  with
 $\alpha \ge 2.3$. For such injected  proton spectra the GZK photons
 reaching us are subdominant at all energies. Details of the fiting procedure
 can be found in Ref.~\cite{min_photon}.

\begin{figure}[ht]
\includegraphics[height=0.5\textwidth,clip=true,angle=270]{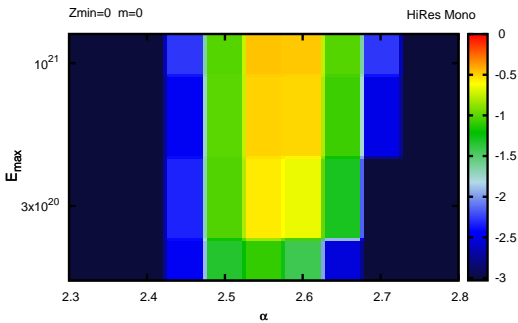}
\caption{Consistency level of the predicted UHECR proton flux
with HiRes data at $E>$2 EeV as function of  $E_{max}$ and $\alpha$  for $m=0$ and
 continuous distribution of sources.  
 Color coded logarithmic $p$-value scale, from
best ($p=1$) to worse ($p$ close to zero).  }
\label{Fig_proton_E_alpha}
\end{figure}

\begin{figure}[ht]
\includegraphics[height=0.5\textwidth,clip=true,angle=270]{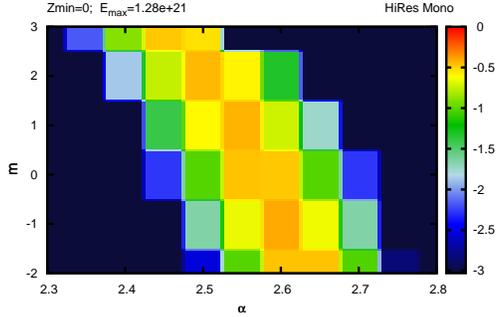}
\caption{Consistency level of the predicted UHECR proton flux
with HiRes data at $E>$2 EeV as function of  $m$ and $\alpha$ for $E_{max}=10^{21}$
eV and  a continuous distribution of sources.  
Color coded logarithmic $p$-value scale, from
best ($p=1$) to worse ($p$ close to zero).  }
\label{Fig_proton_m_alpha}
\end{figure}

In Fig.~\ref{Fig_proton_E_alpha}  and  Fig.~\ref{Fig_proton_m_alpha}
we show the logarithm of the $p$-value in a color coded scale, from
best ($p=1$) to worse ($p$ close to zero), which measures the
consistency level of the predicted UHECR proton flux  with the HiRes
data, for different parameter ranges.

We can see from the figures that  fitting the
UHECR data at  $2$ EeV and above, requires the initial proton spectrum to
be relatively hard, with $\alpha=2.50-2.65$ in
Eq.(\ref{proton_flux}). Fig.~\ref{Fig_proton_E_alpha} shows that this
range does not depend strongly on  $E_{max}$ for a continuous
distribution of sources. In Ref.~\cite{min_photon} it is shown that 
 if instead there are no sources within a distance of
$50$ Mpc, i.e. if $z_{min}=0.01$, the HiRes observed spectrum is not
fitted  as well anymore, and a relatively high maximum energy $E =
10^{21}$ eV is required for a reasonable fit.

The low energy  part of the predicted  spectrum depends mostly on the
power law index $\alpha$ and source evolution index $m$.  In
Fig.~\ref{Fig_proton_m_alpha} we show the goodness of fit $p$-value  as function
of  $m$ and $\alpha$ for $E_{max}=10^{21}$ eV 
for $z_{min}=0$.
This figure cleary shows the degeneracy between the  parameters $m$ and
$\alpha$: as $m$ increases from $-2$ to 3 the value of $\alpha$ of the
best fits decreases from $\simeq 2.6-2.7$ to $\simeq 2.4-2.5$.

Let us  now discuss the secondary photon fluxes. The main
difference between  the minimal model  we are concentrating on here and
other models (see Ref.~\cite{gzk_photon}) is that in the minimal model
one fits the UHECR data with  extragalactic protons from
low  energies $E> 2$ EeV, what requires  a hard
spectrum with  index $\alpha>2.4$ 
(see Figs.~\ref{Fig_proton_E_alpha} and \ref{Fig_proton_m_alpha}).
In this case, as mentioned above,  the GZK photon flux is always sub-dominant, at all
energies.

\begin{figure}[ht]
\begin{center}
\includegraphics[height=0.45\textwidth,clip=true,angle=270]{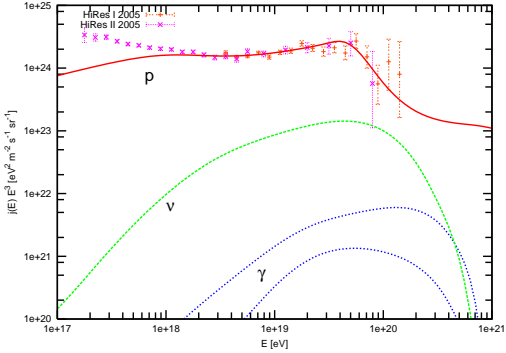}
\end{center}
\caption[...]{Proton, GZK photon and  cosmogenic neutrino spectra for the model with
$m=0$, $z_{min}=0$,  $E_{max}=10^{21}$ eV  and $\alpha=2.55$.  The
upper photon line is for minimal radio background and
$B_{EGMF}=10^{-11}$ G,  while the lower photon line for maximal  radio
background $B_{EGMF}=10^{-9}$ G.}
\label{Fig_photon_spectrum}
\end{figure}

As an example, in   Fig.~\ref{Fig_photon_spectrum}  we show the
possible range of GZK photon fluxes for the 
same proton spectrum. 
Here we do not deal with neutrinos in any detail, but just to compare
the photon and neutrino fluxes produced in the same GZK processes, in
Fig.~\ref{Fig_photon_spectrum} we also plotted the cosmogenic neutrino
flux per flavor for the same model.  Even if the neutrino flux is much
higher than the photon flux, its detection may be even more difficult
due to the strongly reduced probability of neutrinos to produce
air-showers.

In Fig.~\ref{Fig_photon_spectrum} one can see that the best energy
range to find  GZK photons is $E=5-20$ EeV.  At higher energies,  the
small event statistics  will not  allow to find a 1\% fraction of
photons in the UHECR flux, while  at lower energies the photon fraction is
 strongly reduced.

\begin{figure}[ht]
\begin{center}
\includegraphics[height=0.45\textwidth,clip=true,angle=270]{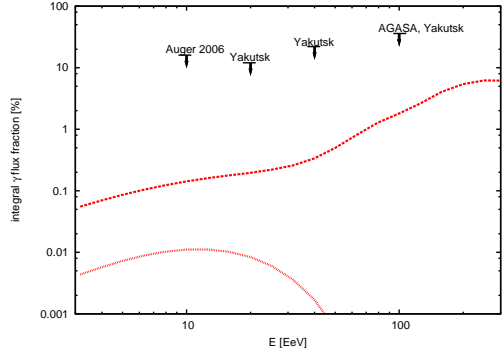}
\end{center}
\caption[...]{Maximum and minimum GZK photon fractions  given  in
percentages  of the integrated flux above  the energy $E$ as function
$E$ for maximum source proton energy $E_{max}=10^{21}$ eV. Present
limits on photon  fraction from Auger~\cite{AugerLimit},
Yakutsk~\cite{Yakutsk_2007} and combined AGASA/Yakutsk
\cite{AgasaYakutskLimit}  data are also shown.}
\label{Fig_photon_E_cut}
\end{figure}

In Fig.~\ref{Fig_photon_E_cut} we show the GZK photon
fraction given in percentage of the integrated UHECR flux above  the
energy $E$ as function of $E$,  for the whole parameter space we consider
(i.e. maximum source proton energy $1.6 \times 10^{20} eV \leq
E_{max}\leq 1.28 \times 10^{21}$ eV,  source evolution parameter  $-2
\leq m \leq 3$, power law index  $2.3 \leq \alpha \leq 2.9$ and
minimum redshift of the sources $0 \leq z_{min} \leq 0.01$).  Present
limits on the  photon fraction from Auger~\cite{AugerLimit},
Yakutsk~\cite{Yakutsk_2007} and combined AGASA/Yakutsk~\cite{AgasaYakutskLimit} 
data are also shown in the figure.  It is
clear that, contrary to the case of top-down models (which are
restricted already by present bounds on the GZK-photon
fraction~\cite{gzk_photon}) the  present limits are well above the
expected the GZK photon fraction in the minimal UHECR model by a
factor  of 10  to 100 depending on the energy (see
Fig.~\ref{Fig_photon_E_cut}). The detection of GZK photons in this
model will remain as a task for the future.

We find that the expected  photon fraction of the integrated flux above
$E=10$ EeV in the minimal UHECR models, is $10^{-4}$ to $10^{-3}$
 independently of the unknown

The  South site of the Pierre Auger Observatory after several years of
data taking will probably be able to reach a photon fraction
sensitivity of the order of $10^{-3}$ in the integrated flux close to
$E=10$ EeV. 
 As can be seen in Fig.~\ref{Fig_photon_E_cut} this is the
level of the largest GZK photon fraction expected in the minimal UHECR
model.
 Larger future observatories like Auger North plus South 
\cite{Auger_North} and EUSO~\cite{EUSO} could probe lower photon fractions if they
are able to collect statistics a  factor of 5-10 larger 
than Auger South and have  thresholds around $1-2 \times 10^{19}$ eV
(provided these experiments
are sensitive to photon primaries).

We have assumed that  the sources emit only protons, however our predictions
for GZK photon fractions  shown in Fig.~\ref{Fig_photon_E_cut}  would
not change  too much if nuclei primaries were present too,  as assumed
in  the so called ``mixed models"~\cite{mixed_model}. The reason is
that even in mixed models, primary protons dominate the UHECR flux
at high energies  $E>50$ EeV, i.e. in the energy  region where the primary
protons produce secondary GZK photons.

As a final remark let us mention that even if the GZK photon fluxes
considered here are very small, much larger fluxes are possible in
more general models, which are not restricted  by the condition that
all the UHECR spectrum from energies $2\times 10^{18}$ eV to the
largest is explained  with extragalactic protons~\cite{gzk_photon}.

%
%
%
%

\end{document}